\documentclass[aps,prl,twocolumn,groupedaddress]{revtex4}

\usepackage{graphicx}
\usepackage{dcolumn}
\usepackage{bm}
\usepackage{amsmath}

\begin{document}


\title{Determination of Spin Contributed Knight Shift of $^{59}\mathrm{Co}$ NMR in $\mathrm{Na_{0.35}CoO_{2} \cdot 1.3H_{2}O}$}


\author{ Chishiro Michioka }
\author{Masaki Kato}
\author{Kazuyoshi Yoshimura}
\affiliation{Department of Chemistry, Graduate School of 
Science, Kyoto University, Kyoto 606-8502, Japan}
\author{Kazunori Takada}
\author{Hiroya Sakurai}
\author{Eiji Takayama-Muromachi}
\author{Takayoshi Sasaki}
\affiliation{Advanced Materials Laboratory, National Institute
for Materials Science, Namiki 1-1, Tsukuba, Ibaraki 305-0044, Japan}


\date{\today}

\begin{abstract}

 The spin contributed Knight shift $^{59}K_{s}$ of $^{59}\mathrm{Co}$ in the triangular lattice superconductor, $\mathrm{Na_{0.35}CoO_{2} \cdot 1.3H_{2}O}$, has been estimated by NMR experiments in an oriented 2-dimensional powder sample. The Knight shift in the paramagnetic state was estimated by $\delta\nu/(\gamma\mathrm{_{N}}H_{i})$ vs. $(\gamma\mathrm{_{N}}H_{i})^{-2}$ plot at various temperatures. A Knight shift vs. magnetic susceptibility diagram revealed that at least 0.3 $\%$ for $K_{x}$ and 0.1 $\%$ for $K_{y}$ are attributed to the spin contribution. Below $T\mathrm{_{c}}$, $\delta\nu/(\gamma\mathrm{_{N}}H_{i})$ vs. $(\gamma\mathrm{_{N}}H_{i})^{-2}$ plot does not give unique Knight shift because of the diamagnetic effect and/or the other ones. We concluded that the invariant behavior at about 7 T (between $H\mathrm{_{c1}}$ and $H\mathrm{_{c2}}$) below $T\mathrm{_{c}}$ possibly indicates the nature of the spin-triplet superconductivity.

\end{abstract}


\pacs{Valid PACS appear here}


\maketitle

\title{Determination of The Spin Contributed Knight Shift in  $\mathrm{Na_{0.35}CoO_{2} \cdot 1.3H_{2}O}$}

\author{Chishiro Michioka}
\author{Masaki Kato}
\author{Kazuyoshi Yoshimura}
\affiliation{Department of Chemistry, Graduate School of 
Science, Kyoto University, Kyoto 606-8502, Japan}
\author{Kazunori Takada}
\author{Hiroya Sakurai}
\author{Eiji Takayama-Muromachi}
\author{Takayoshi Sasaki}
\affiliation{Advanced Materials Laboratory, National Institute
for Materials Science, Namiki 1-1, Tsukuba, Ibaraki 305-0044, Japan}

\date{\today}

 Superconductivity in $\mathrm{Na_{0.35}CoO_{2} \cdot 1.3H_{2}O}$ with $T_{c} \sim $ 5 K was discovered recently. \cite{takada} In this compound, the triangular configuration in $\mathrm{CoO_{2}}$ plane has a possibility of an unconventional superconductivity in the novel quantum state. Indeed, many theoretical models for the $\mathrm{CoO_{2}}$ plane have been investigated, and pointed out a possibility that this superconductivity originates in a marvelous superconducting mechanism. For example, Tanaka $et$ $al$. indicated that the spin-triplet superconductivity may be realized with similar pairing process to $\mathrm{Sr_{2}RuO_{4}}$ since the hexagonal structure in $\mathrm{Na_{0.35}CoO_{2} \cdot 1.3H_{2}O}$ gives a stronger spin-triplet pairing tendency. \cite{tanaka} On the other hand, some groups studied a single band t-J model as an appropriate model to understand the low energy electronic phenomena resulting that the superconductivity in $\mathrm{Na_{0.35}CoO_{2} \cdot 1.3H_{2}O}$ occurs related to the resonating-valence-bond (RVB) state. \cite{baskaran1, kumar, wang, ogata} Furthermore, Ikeda $et$ $al$. investigated superconducting instabilities on a 2D triangular lattice with the repulsive Hubbard model. Their result shows that the $f$-wave pairing with the triplet superconductivity is stable in the wide range of the phase diagram. \cite{ikeda} Quite recently, Kuroki $et$ $al$. introduced a single band effective model with taking into account the pocket-like Fermi surfaces along with the van Hove singularity near the K point, and showed that the large density of states near the Fermi level gives rise to ferromagnetic spin fluctuations, leading to a possible realization of the $f$-wave superconductivity due to the disconnectivity of the Fermi surfaces near the $\Gamma$ point. \cite{kuroki1, kuroki2, kuroki3, kuroki4}

In order to clarify the superconducting mechanism in $\mathrm{Na_{0.35}CoO_{2} \cdot 1.3H_{2}O}$, it is very important to investigate the Cooper-pair symmetry. Nuclear magnetic resonance (NMR) is known as the powerful probe to decide the pairing symmetry in the superconducting state. The nuclear quadrupole resonance (NQR) studies of $\mathrm{^{59}Co}$ revealed that there is no coherence peak in the temperature dependence of the nuclear spin-lattice relaxation rate $1/T_{1}$ suggesting that the unconventional superconductiving state with line-node gap. \cite{fujimoto, ishida} The present authors studied $1/T_{1}$ up to 7.12 T by $\mathrm{^{59}Co}$ NMR, and found that the transition temperature at 7.12 T is about 4.4 K. \cite{waki} This anomalously small decrease of $T_{c}$ suggests large $H\mathrm{_{c2}}$, which is consistent with macroscopic properties. \cite{sakurai} Moreover, field-swept NMR spectra suggested that the resonance magnetic fields in constant frequencies do not change below $T\mathrm{_{c}}$ at about H = 7 T which is far below $H\mathrm{_{c2}}$. This means that the Knight shift is invariant below $T\mathrm{_{c}}$, which results in the spin-triplet superconductivity in this system. However, this result is inconsistent with those reported by Kobayashi $et$ $al$., which suggested spin-singlet $s$-wave superconductivity. \cite{sato} It is generally difficult to determine NMR parameters including Knight shift uniquely from a field-swept spectrum taken at only one constant frequency. In this paper, we first present the temperature dependence of the Knight shift above $T\mathrm{_{c}}$ and show how to estimate its spin part, which is very important for the superconducting state as well. Next we discuss the spin contributed Knight shift below $T\mathrm{_{c}}$, in which the diamagnetic effect would be considered. 

 The powder sample of $\mathrm{Na_{0.35}CoO_{2} \cdot 1.3H_{2}O}$ was prepared by the oxidation process from the mother compound, $\mathrm{Na_{0.7}CoO_{2}}$. \cite{takada} The superconducting transition temperature ($T\mathrm{_{c}}$) is estimated by the macroscopic magnetization using SQUID magnetometer. $\mathrm{^{59}}$Co NMR spectra were measured by the spin-echo method with a phase coherent-type pulsed spectrometer. The powder sample was oriented and fixed by using an organic solvent Hexane under the condition of $H$ = 8 T. \cite{condition} The X-ray diffraction measurement indicated that $c$-axes of powders were oriented perpendicular to the external magnetic field $H$, and random orientation in the basal plane was obtained along $H$. Therefore our NMR measurements were done on the two-dimensional (2D) powder sample in the condition with the magnetic field being perpendicular to the $c$-axis. 

 From the following two reasons, we took care of the following experimental condition that once the sample has cooled, and then the temperature of the chamber was kept below 50 K, until all the experiments had been finished. The first reason is that the specific resonance magnetic field corresponding to a peak on the field-swept NMR spectrum should be changed when the condition of the sample orientation was changed. In other words, a degree of the orientation should not be changed during whole measurements. The second is that the change of the site and its occupancy of the water molecules within the sample may change the electronic state as well as the quadrupole frequency $\nu\mathrm{_Q}$ because the increase of the temperature enables water molecules to move. In this case, the intrinsic Knight shift cannot be obtained correctly. 

\begin{figure}
\begin{center}
\includegraphics[width=0.47\linewidth]{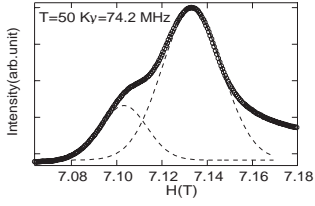}
\end{center}
\caption{\label{fig:peak} Field-swept NMR spectrum of $\mathrm{Na_{0.35}CoO_{2} \cdot 1.3H_{2}O}$ at T = 50 K and $\nu$ = 74.2 MHz. Two peaks, which are decomposed by the Gaussian fitting indicated by the dashed lines, correspond to the central resonance of $I_{z} = -1/2 \leftrightarrow 1/2$ transition. The analysis of the anisotropic Knight shifts $K\mathrm{_{x}}$ and $K\mathrm{_{y}}$ were done using the low-field smaller peak and the high-field larger peak, respectively (see the text). }
\end{figure}

Figure \ref{fig:peak} shows the typical central peak of the field-swept spectrum, which corresponds to the central transition ($I_{z} = -1/2 \leftrightarrow 1/2$) at T = 50 K and $\nu$ = 74.2 MHz. The whole spectrum and its assignment are explained in detail in a different paper. \cite{waki} In this paper, we carefully discuss the resonance field depending upon the resonance frequency in order to determine the value of Knight shift uniquely.

\begin{figure}
\begin{center}
\includegraphics[width=0.8\linewidth]{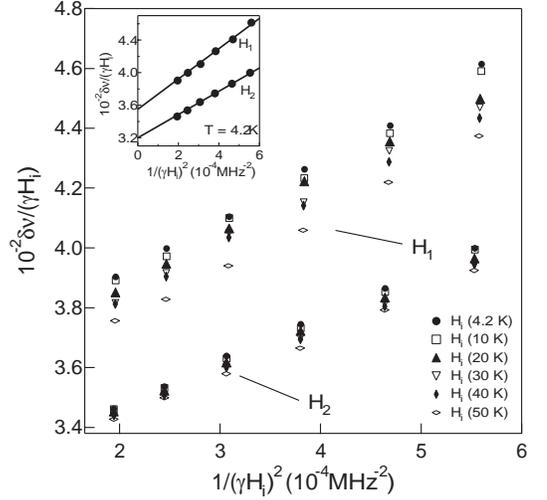}
\end{center}
\caption{\label{fig:aboveTc} $\delta\nu/(\gamma\mathrm{_{N}}H_{i})$ vs. $(\gamma\mathrm{_{N}}H_{i})^{-2}$ ($i$ = 1, 2) plot at various temperatures above $T\mathrm{_{c}}$ and 4.2 K (just below $T\mathrm{_{c}}$) in the frequency range 44.2 - 74.4 MHz. $H\mathrm{_{1}}$ and $H\mathrm{_{2}}$ are the resonance fields at which the $\mathrm{^{59}Co}$ NMR spectrum with 2D-powder pattern of the central transition show two singular points. Inset shows data at 4.2 K with the fitted lines using the NMR parameters, $\nu_{Q}$ = 4.15 MHz, $\eta$ = 0.21, $K_{x}$ = 3.56 \% and $K_{y}$ = 3.21 \%. }
\end{figure}

\begin{figure}
\begin{center}
\includegraphics[width=0.8\linewidth]{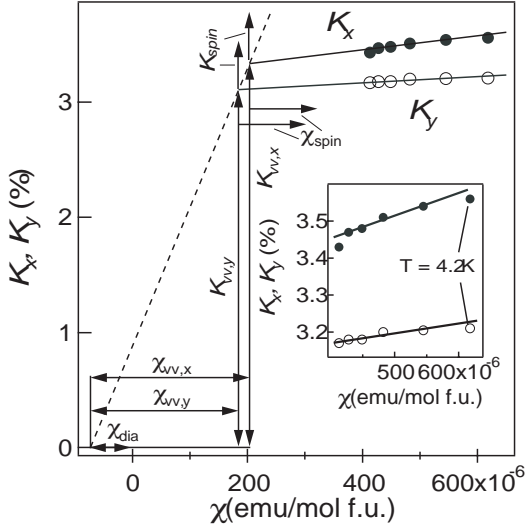}
\end{center}
\caption{\label{fig:Kchi} Knight shift vs. susceptibility diagram in $\mathrm{Na_{0.35}CoO_{2} \cdot 1.3H_{2}O}$. The data at 4.2 K (below $T\mathrm{_{c}}$) is shown for comparison but not used for the line fitting. Inset shows the expansion. The numerical details for constructing the diagram are given in the text. }
\end{figure}

In this process, we assume that principal axes of the Knight shift tensor coincide with those of the electric-field gradient (EFG) tensor. Under this assumption, since we measured on a 2D-powder sample, we could not estimate the $Z$ component of the Knight shift. In generally, five specific resonance frequencies of the central transition appear in a constant field on the powder sample for the case of $\eta < \frac{1}{3}$, where $\eta$ is the asymmetric parameter of the EFG tensor defined as $\eta = |V_{XX}-V_{YY}|/|V_{ZZ}|$, in which $V_{ii}$ ($ii$ = $XX$, $YY$, $ZZ$) are the principal components of the EFG tensor with the relation, $|V_{XX}| \le |V_{YY}| \le |V_{ZZ}|$. \cite{text} That is, five specific fields ($H\mathrm{_{1}}$-$H\mathrm{_{5}}$) appear as two peaks, one step and two edges in a constant frequency. In the case of the 2D-powder, two specific resonance fields ($H\mathrm{_{1}}$ and $H\mathrm{_{2}}$) appear as singular points, and the rest specific fields disappear because of the 2D orientation. Hence the fields, at which the field-swept spectrum shows peaks as shown in Fig. \ref{fig:peak}, should correspond to $H\mathrm{_{1}}$ and $H\mathrm{_{2}}$. If the quadrupole effect is treated by the second-order perturbation, the linear relation between $\delta\nu_{i}$/$\gamma\mathrm{_{N}}H_{i}$ and $(\gamma\mathrm{_{N}}H_{i})^{-2}$ can be obtained as follows, 
\begin{equation}
\frac{\delta\nu_{i}}{\gamma\mathrm{_{N}}H_{i}} = K_{j} + \frac{C_{i}}{(1+K_{j})(\gamma\mathrm{_{N}}H_{i})^{2}},
\label{eq:A}
\end{equation}

\begin{equation}
\delta\nu = \nu_{0} - \gamma_\mathrm{{N}}H_{i}
\label{eq:B}
\end{equation}

\begin{equation}
C_{1} = \frac{R(3+\eta)^{2}}{144},  C_{2} = \frac{R(3-\eta)^{2}}{144},
\label{eq:C}
\end{equation}

\begin{equation}
R = \nu_{Q}^{2}[I(I+1)-\frac{3}{4}],
\label{eq:D}
\end{equation}
where the values with $j = X, Y$ correspond to those with $i$ = 1, 2, respectively, $\gamma\mathrm{_{N}}$ is the nuclear gyromagnetic ratio (10.054 MHz/T for $\mathrm{^{59}Co}$ with the nuclear spin of $I$ = 7/2), $\nu_{0}$ the resonance frequency and $\nu_{Q}$ the quadrupole frequency. Figure \ref{fig:aboveTc} shows $\delta\nu/(\gamma\mathrm{_{N}}H_{i})$ vs. $(\gamma\mathrm{_{N}}H_{i})^{-2}$ ($i$ = 1, 2) plot of $\mathrm{Na_{0.35}CoO_{2} \cdot 1.3H_{2}O}$ above $T\mathrm{_{c}}$ and 4.2 K in the frequency range 44.2 - 74.4 MHz. Here, we can estimate anisotropic Knight shift uniquely as an interception of the $\delta\nu_{i}$/$\gamma\mathrm{_{N}}H_{i}$ axis in this plot under the assumption that the Knight shift is independent of $H$. The resonances of $H_{1}$ and $H_{2}$ show clear linearity in this plot, and $K_{x}$ and $K_{y}$ (interceptions of the $\delta\nu_{i}$/$\gamma\mathrm{_{N}}H_{i}$ axis) change with $T$ while the slopes are almost independent of $T$, which is consistent with the fact that the NQR frequency is almost constant in these temperatures, because the slope $C_{i}$ is determined mainly by quadrupole parameters. \cite{Ishida2} The inset in Fig. \ref{fig:aboveTc} shows the fitting using Eqs. (\ref{eq:A})-(\ref{eq:D}) at 4.2 K. Slopes of the two lines depending on Eq. (\ref{eq:B}) give parameters of the quadrupole interaction, $\nu_{Q}$ = 4.15MHz and $\eta$ = 0.21, and these values are consistent with the resonance fields of the steps of the first, second and third satellites appeared in a whole spectrum. \cite{waki} As a result, $K_{x}$ and $K_{y}$ decrease with increasing temperature, which have scaling relationship with the magnetic susceptibility. Therefore we can estimate the contribution of each interaction for the Knight shift from the Knight shift vs. susceptibility ($K-\chi$) plot shown in Fig. \ref{fig:Kchi}. 

In terms of the corresponding susceptibilities three contributions to the Knight shift can be expressed as 
\begin{equation}
K(T) = K_{spin}(T) + K_{ce} + K_{VV},
\label{eq:E}
\end{equation}
where the temperature-depending term $K_{spin}(T)$ and the temperature independent term $K_{ce}$ are attributed to the spin susceptibilities of $d$ spins and conduction electrons or holes, respectively, and the last term is due to the orbital susceptibility of $d$ electrons. The diamagnetic susceptibilities of core electrons can be estimated from relativistic Hartree-Fock calculations, in which $\chi_{dia}^\mathrm{{H}}$ = $ - 0.4 \times 10^{-9}$ $\mu\mathrm{_{B}/atom}$, $\chi_{dia}^\mathrm{{O}}$ = $ - 1.6 \times 10^{-9}$ $\mu\mathrm{_{B}/atom}$, $\chi_{dia}^\mathrm{{Na}}$ = $ - 3.8 \times 10^{-9}$ $\mu\mathrm{_{B}/atom}$ and $\chi_{dia}^\mathrm{{Co}}$ = $ - 5.5 \times 10^{-9}$ $\mu\mathrm{_{B}/atom}$. \cite{Mendelsohn} The total diamagnetic susceptibility of $\mathrm{Na_{0.35}CoO_{2} \cdot 1.3H_{2}O}$ was described in Fig. \ref{fig:Kchi} as $\chi_{dia}$. The hyperfine coupling constant for the $d$ orbital moment, $A_{hf}^{VV} = \frac{K_{VV}}{\chi_{VV}} = 6.7\times10^{2}$ kOe/$\mu_{B}$ was estimated as a product of $2/\langle r^{-3}\rangle = 9.05\times10^{25}\mathrm{cm}^{-3}$ for $\mathrm{Co^{3+}}$ and the reduction factor of 0.8 for the metallic state compared with the free atom. Because the strict band structure of $\mathrm{Na_{0.35}CoO_{2} \cdot 1.3H_{2}O}$ is unknown, $K_{ce}$ cannot be estimated correctly. We ignore the small contribution by the conduction electrons because the contribution is much smaller than the uncertainty of the reduction factor. The coupling constants for the temperature dependent Knight shift components, $A_{hf}^{x}$ and $A_{hf}^{y}$ are estimated as 34 kOe/$\mu_{B}$ and 14.7 kOe/$\mu_{B}$ by the least-square fitting shown as solid lines in Fig. \ref{fig:Kchi}. Then $\chi_{VV,x}$ and $\chi_{VV,y}$ are estimated as $2.0\times10^{-6}$ and $1.8\times10^{-6}$ emu/mol f.u., respectively, and $K_{VV,x}$ and $K_{VV,y}$, 3.3 and 3.1 $\%$, respectively. Therefore spin contributed Knight shifts are at least 0.3 $\%$ for $K_{x}$ and 0.1 $\%$ for $K_{y}$.

\begin{figure}
\begin{center}
\includegraphics[width=0.68\linewidth]{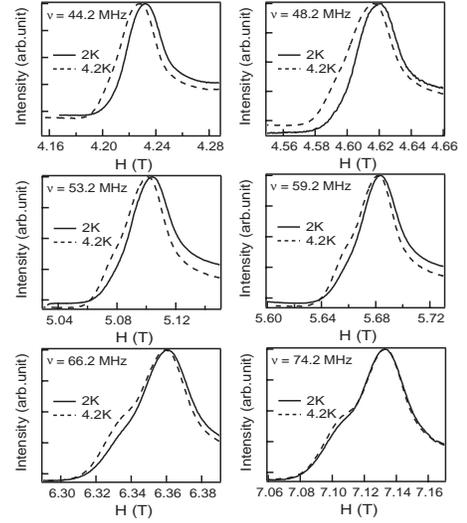}
\end{center}
\caption{\label{fig:spectra} Field-swept 2D-powder spectra for the central transition at 2.0 and 4.2 K in several constant frequencies $\nu$.}
\end{figure}

\begin{figure}
\begin{center}
\includegraphics[width=0.8\linewidth]{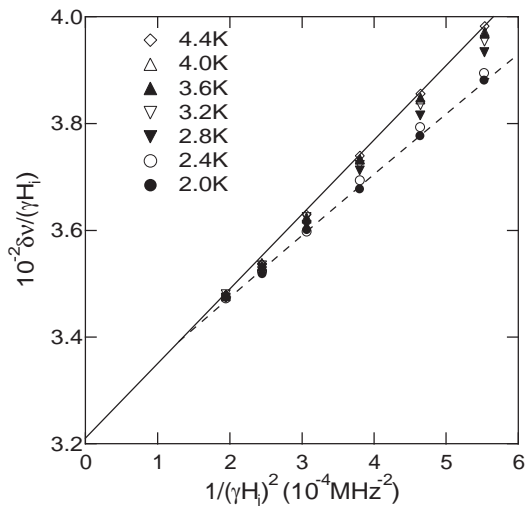}
\end{center}
\caption{\label{fig:belowTc} $\mathrm{\delta\nu/(\gamma\mathrm{_{N}}H_{2})}$ vs. $(\gamma\mathrm{_{N}}H_{2})^{-2}$ plot at various temperatures below $T\mathrm{_{c}}$. Lines in the figure are guide for eyes. The linear correlation which is characteristic above $T\mathrm{_{c}}$ disappears below $T\mathrm{_{c}}$. This is not due to the change of $\nu_{Q}$ or $\eta$ but due to the effect of the diamagnetization.}
\end{figure}

Next we discuss about the temperature dependence of the Knight shift below $T\mathrm{_{c}}$. Figure \ref{fig:spectra} shows the central peaks in the field-swept spectra at 2.0 and 4.2 K at several constant frequencies (44.2 - 74.2 MHz). While the resonance fields are almost the same between 2.0 and 4.2 K at $\nu$ = 74.2 MHz, it changes drastically at $\nu$ = 44.2 MHz. The differences of the resonance fields between 2.0 and 4.2 K increase gradually with decreasing the resonance frequency. In the previous paper, we showed that the field $H$ = 7 T is obviously smaller than $H\mathrm{_{c2}}$ from the temperature dependence of $1/T_{1}T$, in which the marked decrease of $1/T_{1}T$ due to the decrease of the density of states around Fermi surface in the superconducting state appears below 4.4 K. As well as in the paramagnetic state, we discuss about the $\delta\nu/(\gamma\mathrm{_{N}}H_{2})$ vs. $(\gamma\mathrm{_{N}}H_{2})^{-2}$ plot in the superconducting state shown in Fig. \ref{fig:belowTc}. We cannot detect $H_{1}$ because of the difficulty in decomposing these spectra especially at lower frequencies. The plot is almost linear at 4.4 K as similar to those above $T\mathrm{_{c}}$, in which the slope is consistent with the fact that $\nu_{Q}$ and $\eta$ do not change in these temperatures. Since $\nu_{Q}$ and $\eta$ do not change also below $T\mathrm{_{c}}$, it makes no physical sense to fit the data by the line having different slope below $T\mathrm{_{c}}$. In this case, if one estimates the Knight shift from the resonance field taken at only one constant frequency, the intrinsic value of the Knight shift should not be estimated accurately. Two explanations can be adopted about the anomalous behavior below $T\mathrm{_{c}}$. One is that the Knight shift depends on $H$ and the other is the diamagnetic or the other unexpected effects. The possibility of the former case is, however, excluded because the intrinsic Knight shift in the superconducting state should not change even if the resonance field on appearance changes due to the distribution between Knight shift of the intrinsic superconducting and normal state, in which $H$ penetrates. The fact that the difference of the resonance fields between 2.0 and 4.4 K is decreasing with increasing the resonance frequency coincides qualitatively to the diamagnetic effects. In addition to this fact, since the resonance field is almost the same between at 2.0 and 4.2 K at $\nu$ = 74.2 MHz, the intrinsic behavior of the Knight shift in the superconducting state is invariant, suggesting that the symmetry of the Cooper pair of the superconducting state in $\mathrm{Na_{0.35}CoO_{2} \cdot 1.3H_{2}O}$ is of the $p$ or $f$ wave with the triplet one. The result is consistent with the $\mu$SR study on $\mathrm{Na_{0.35}CoO_{2} \cdot 1.3H_{2}O}$. \cite{higemoto}

 We estimated the spin contributed Knight shift of $\mathrm{Na_{0.35}CoO_{2} \cdot 1.3H_{2}O}$. In the paramagnetic region, the spin contributed Knight shifts are at least 0.3 and 0.1 $\%$ for $K_{x}$ and $K_{y}$. In the superconducting state, the Knight shift is not estimated correctly, however the intrinsic Knight shift should not change at the external field up to at least 7 T. From these results, we concluded that the $p$ or $f$ wave pairing state with triplet spin symmetry may be most preferable in the superconducting state of $\mathrm{Na_{0.35}CoO_{2} \cdot 1.3H_{2}O}$.

\begin{acknowledgments}
cussions. This study was supported by a Grant-in-Aid on priority area 'Novel Quantum Phenomena in Transition Metal Oxides', from Ministry of Education, Science, Sports and Culture (12046241), and also supported by a Grant-in-Aid Scientific Research of Japan Society for Promotion of Science (12440195, 12874038)
\end{acknowledgments}


\begin{references}
\bibitem{condition} In our NMR measurement, Hexane was the optimal as an agent which fix the sample because it does not absorb water molecules.
\end{references}
\end{document}